# Photocatalytic activity enhancement by addition of lanthanum into the BiFeO$_3$ structure and the effect of synthesis method

*Hamed Maleki*[*]

Faculty of Physics, Shahid Bahonar University of Kerman, Kerman, Iran

e-mail address: hamed.maleki@uk.ac.ir

**Abstract:** In this paper, the photocatalytic activity of multiferroics BiFeO$_3$ (BFO) and Bi$_{0.8}$La$_{0.2}$FeO$_3$ (BLFO) nanocrystals with two different morphologies which were synthesized by two different sol-gel (SG) and hydrothermal (HT) methods have been studied. All the obtained samples were characterized using X-ray diffractometer, Fourier transform infrared spectroscopy, transmission electron microscopy, UV-vis spectroscopy and vibrating sample magnetometer. Differential thermal analysis (DTA) measurements were probed ferroelectric-paraelectric first-order phase transition (T$_C$) for all samples. Addition of lanthanum decreases the electric phase transition. For photocatalyst application of bismuth ferrite, adsorption potential of nanoparticles for methylene blue (MB) organic dye was evaluated. The doping of La in the BFO structure enhanced the photocatalytic activity and about 71% degradation of MB dye was obtained under visible irradiation. The magnetic and ferroelectric properties of BLFO nanoparticles improve compared to the undoped BiFeO$_3$ nanoparticles. The non-saturation at high applied magnetic field for as-prepared samples by HT is related to the size and shape of products. This work not only presents an effect of lanthanum substitution into the bismuth ferrite structure on the physical properties of BFO, but also compares the synthesis method and its influence on the photocatalytic activity and multiferroics properties of all nanopowders.

## 1. Introduction

In order to generate hydrogen, an environmentally friendly process has been offered to modern society through photocatalytic degradation of pollutants and photocatalytic water splitting using solar energy [1–3]. Multiferroic materials have recently drawn scientist's attention due to their vast applications as a result of their photocatalysis and photovoltaics properties. The narrow energy band gap and ferroelectric properties lead to high absorption of light in the visible region [4–7]. Bismuth ferrite, which simultaneously shows ferroelectric and ferromagnetic behavior at and above room temperature (RT), has drawn attention for two decades. This increasing interest in bismuth ferrite as a noble material is due to its potential application in multifunctional devices, data storage, sensors and photovoltaic technologies [8–18]. This matter has the anti(ferro)-magnetic to paramagnetic transition at Neel temperature $T_N\sim370\ ^oC$ and ferroelectric-paraelectic first order phase transition at Curie temperature $T_C\sim830\ ^oC$ [19–23].

Literature survey indicates that, with periodicity of 62 nm spin structure in BFO gives a spiral modulation and is a cause of G-type antiferromagnetic ordering of bulk bismuth ferrite [24–26]. Moreover, due to the formation of secondary phase and impurities during synthesis process, BFO has large leakage current density (because of various oxidation states of Fe ion and the existence of oxygen vacancies) [8,27–29]. Many theoretical and experimental studies of bismuth ferrite have been carried out to expand the applications and solve the problems hindering practical usage of BFO [30–32]. To overcome these obstacles, parallel to synthesis of BFO nanoparticles with different methods [11,33–35], it has been reported that A-site or B-site substitution into the structure of bismuth ferrite, is the most effective strategy to reduce the impurity phases and enhanced multifferoic properties [36–44].

Apart from these multiferroic properties of $BiFeO_3$, according to the narrow band-gap of BFO (~2.2 eV), bismuth ferrite has been also known as a magnetic photocatalyst at visible light irradiation for water splitting and degradation of organic pollutants [5,7,45]. The weak ferromagnetic nature of $BiFeO_3$ nanoparticles is also used for recovering the catalyst from solution. However, the photocatalytic efficiency of bismuth ferrite is also limited as a result of its low conduction band (compared to $H_2$ or $O_2$) and small surface areas for catalytic reactions [1,46]. In order to enhance photocatalytic ability on the degradation of pollutants, BFO has been synthesized by varying its synthesis method and by varying its compositional parameters such as substitution.

To understand the origin of the influence of synthesis method, as well as the effect of addition of lanthanum into the structure of bismuth ferrite, we have synthesized $BiFeO_3$ and $Bi_{0.8}La_{0.2}FeO_3$ nanoceramics through the sol-gel and hydrothermal methods. The synthesis was done to find a deep understanding about their effects on the multiferroic properties and photocatalytic activities of BiFeO$_3$ nanoparticles.

## 2. Experimental method

### 2.1 Sol-gel preparation of pure and La-doped BiFeO$_3$

BiFeO$_3$ and Bi$_{0.8}$La$_{0.2}$FeO$_3$ nanoceramics were synthesized via sol-gel (SG) process by using Bi(NO$_3$)$_3$.5H$_2$O, Fe(NO$_3$)$_3$.9H$_2$O and La(NO$_3$)$_3$.6H$_2$O (for the case of BLFO) as a starting materials, and deionized water and 2-methoxyethanole as a solvent [47,48]. Stoichiometric amount of bismuth, iron and lanthanum nitrate were completely dissolved in the solvent. Acetic acid was added to the above solution dropwise (pH~1.5) one hour later. A brownish and transparent sol is obtained after 1.5 h of constant stirring. Then the temperature was increased to 90°C and the solution was slowly evaporated and after 3h heating and stirring, the gel is obtained. The precursor powders were dried for 2h at 110°C and were calcined at 650°C for 3h. The as-prepared products synthesized through the sol-gel process were indexed SGBFO and SGBLFO for pure and La-doped bismuth ferrite respectively.

### 2.2 Hydrothermal synthesis of BFO and BLFO

In this method, for synthesis of BiFeO$_3$ (HTBFO) and Bi$_{0.8}$La$_{0.2}$FeO$_3$ (HTBLFO) samples, a stoichiometric amount of bismuth (III) nitrate pentahydrate (99%, Sigma), ferric (III) nitrate nonahydrate (99%, Merck) and lanthanum nitrate hexahydrate (99% Merck for HTBLFO) were dissolved in 10 ml of deionized water. Then 30 ml KOH (4M) was added to the solution dropwise under constant stirring. After 30 min stirring, the solution was transferred into a stainless Teflon-lined autoclave and heated at 220°C for 12h. Then the solution cooled naturally to the RT. The products were separated from the solution by centrifugation, then washed several times with distilled water and ethanol, and finally dried for 2 hours at 110 °C, in order to obtain brown nanopowders.

### 2.3 Characterization

First, the physical properties of pure and La-doped bismuth ferrite is studied. The crystallinity and structural properties of products were investigated by X-ray diffractometer (XRD) on a Philips X'pert, with Cu-Kα radiation (λ=1.54056 Å) and Furrier transform inferared (FTIR) TENSOR27 spectrophotometer. The morphology and distribution of nanocrystals were observed by using transmission electron microscopy (TEM, Leo-912-AB). The thermal treatment and weight loss of the nanoparticles were recorded by differential thermal and thermal gravimetric analysis (TG-DTA, NETZSCH- PC Luxx 409) at a heating rate of 10°C/min from RT up to 1000°C in air. The magnetic properties were measured by a Lake Shore (7410, SAIF) vibrating sample magnetometer (VSM) up to a maximum field of 20 kG. The optical properties of products were examined by UV-vis absorption spectra using Lambda900 spectrophotometer.

In the second part, the photocatalytic activity of all samples were evaluated by the degradation of methylene blue in aqueous solution using 300 W Xenon lamp under visible light irradiation. The starting concentration of MB was chosen 4 mgl$^{-1}$ with dispersing 0.2g BFO or BLFO in aqueous solution of 200 ml. For achieving an adsorption equilibrium of MB on products surface, the aqueous suspension was stirred magnetically for a period of 75 min in the dark prior to irradiation. Then by turning on the lamp the changes of MB concentration were measured. Concentration variations were studied by measuring the absorbance of the solution at 664 nm using a UV-vis spectrophotometer. $C/C_0$ is known as the photocatalytic degradation ratio of MB, which has been investigated in this study. $C_0$ was the starting concentration of MB and C was the concentration of MB at time t.

## 3. Results and discussion
### 3.1 X-ray diffraction investigation

Fig. 1 shows the XRD patterns of BFO and BLFO nanoparticles prepared by sol-gel (Fig. 1 (a) and (b)) and hydrothermal (Fig. 1 (c) and (d)) methods. The recorded diffracted planes match well with the standard card of bismuth ferrite (JCPDS card No. 86-1518) and XRD patterns confirm the single perovskite phase of BiFeO$_3$ with distorted rhombohedral structure (R$_3$C). All patterns show strong and sharp diffraction peaks which indicate well crystallinity of all as-prepared particles. For pure BFO samples a few weak diffraction peaks related to the impurity phases like Bi$_2$Fe$_4$O$_9$ and Bi$_{25}$FeO$_{40}$ are observed. By adding La into the structure of BFO, these peaks are remained for HTBLFO sample, however for SGBLFO nanoparticles the impurity phases disappeared. For nanoparticles prepared by SG, dopant lanthanum changes the position of peaks a little to the smaller value and addition of La into the structure of bismuth ferrite, leads to a phase transformation from rhombohedral to monoclinic structure which is observed in the merging of two major peaks of BFO at 30<2θ<35 into the one peak (Figs.2 (b) and (d)) [47]. On the other hand, for HTBFO and HTBLFO nanoparticles, at a larger angle, the major peaks exhibit no phase transformation. The average crystallite size, has been obtained from the Scherrer formula D= $\frac{K\lambda}{\beta \cos\theta}$ , where *K* is the shape factor that normally measures to be about 0.89, λ is the wavelength of Cu-Kα radiation of the XRD, β is the width of the observed diffraction peak at its half intensity maximum, and θ is the Bragg angle of each peak. The calculated average size for SGBFO and SGBLFO grains are 48 nm and 34 nm respectively.

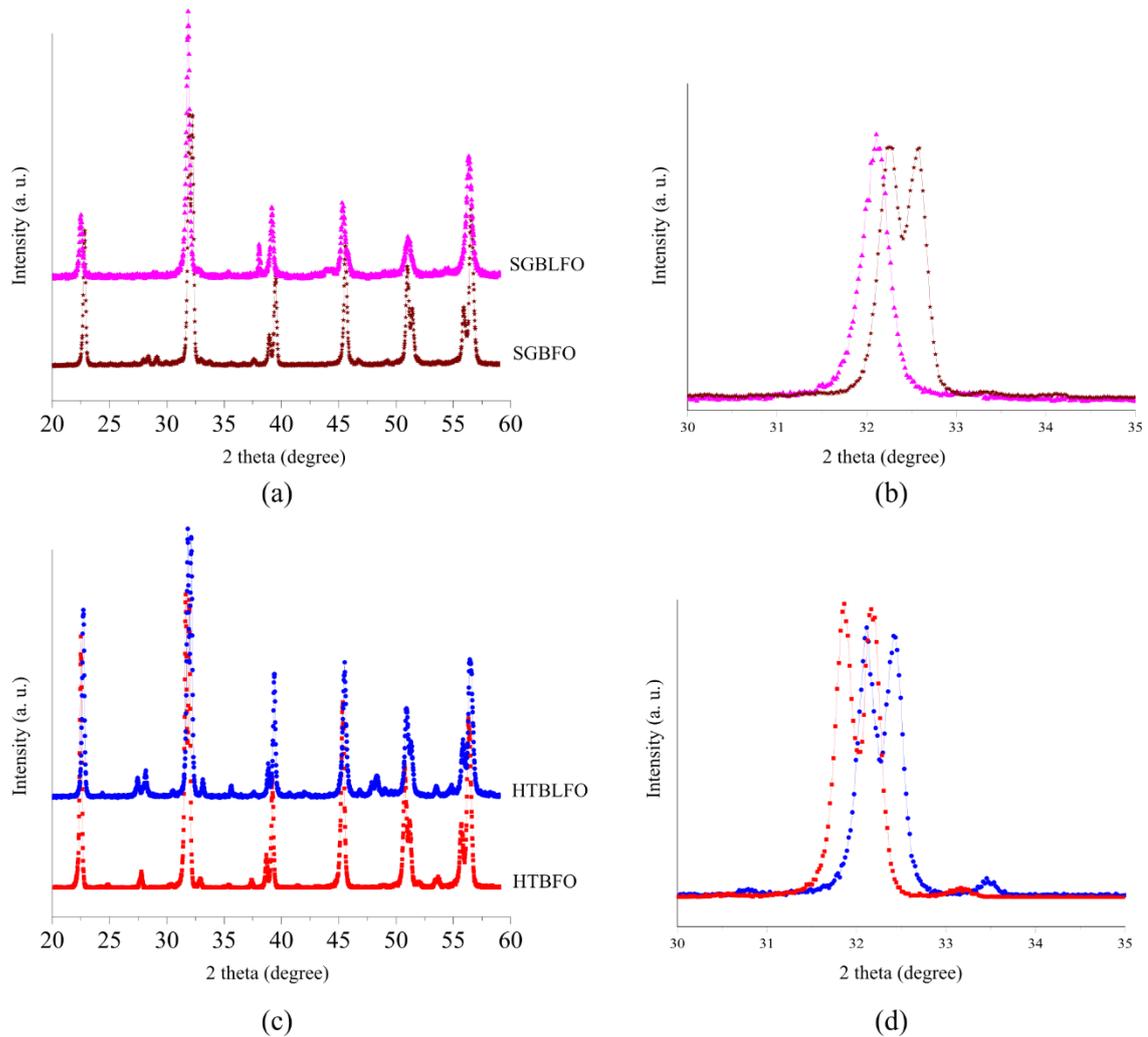

**Fig. 1** (a) XRD patterns of BiFeO$_3$ and Bi$_{0.8}$La$_{0.2}$FeO$_3$ nanoparticles prepared by SG method, (b) same plot in the range of 30<2θ<35, (c) XRD patterns of HTBFO and HTBLFO nanoparticles, (d) same graphs in the range of 30<2θ<35.

**3.2 Furrier transform infrared spectroscopy**

To get further insight into the formation mechanism of BFO nanoparticles, the FTIR spectra of all samples were recorded. Fig. 2 illustrate the FTIR spectra of all as-synthesized nanoceramics in the wavenumber range of 400-4000 cm$^{-1}$. Two major peaks between 400-600 cm$^{-1}$ correspond to the Fe-O stretching and O-Fe-O bending vibrations of perovskite FeO$_6$ groups, which confirms the formation of BFO nanoparticles [11,49,50]. Between 1400-1650 cm$^{-1}$ two peaks are observed which are related to the symmetry bending vibration of C-H or C-

$H_2$ [51]. The broad band at 3300-3700 cm$^{-1}$ is due to the symmetric and anti-symmetric stretching of $H_2O$ and $OH^-$ bond.

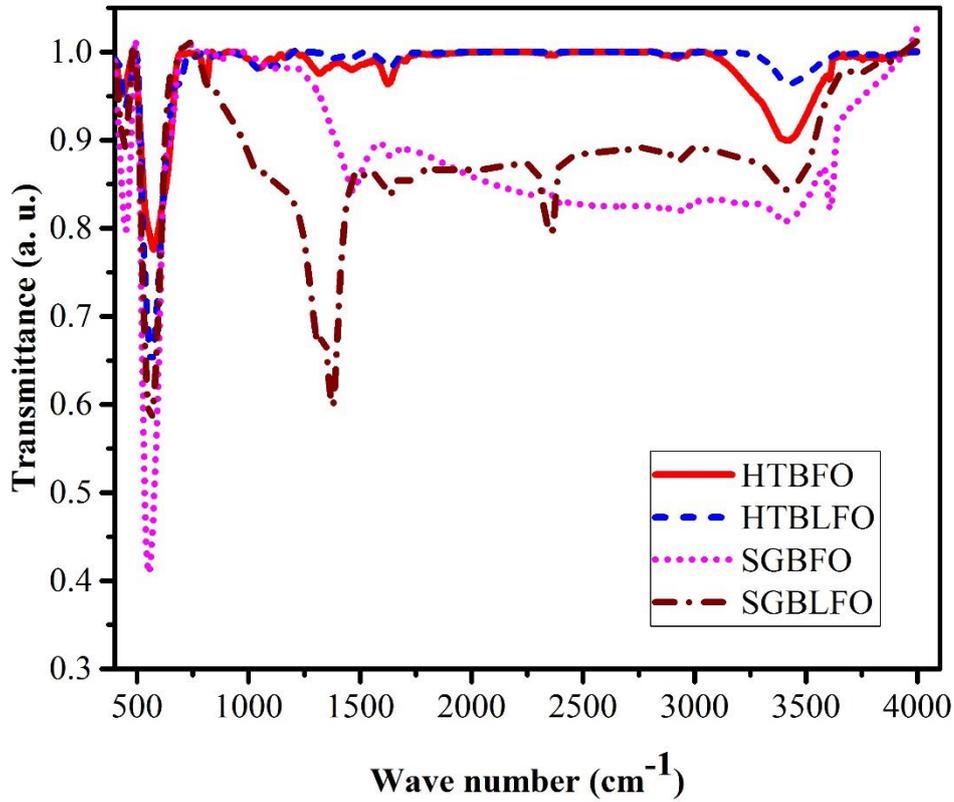

**Fig. 2** FTIR spectra of $BiFeO_3$ and $Bi_{0.8}La_{0.2}FeO_3$ nanoparticles prepared by both SG and HT methods

### 3.3 Transmission electron microscope

TEM images of un-doped and La-doped BFO nanoparticles synthesized by SG and HT methods are presented in Fig. 3. The sol-gel as-prepared nanoparticles which are shown in Fig. 3 (a) and (b), have polyhedral morphology with the average size of 48 nm and 34 nm for SGBFO and SGBLFO nanoparticles respectively. TEM images of HTBFO and HTBLFO nanoceramics are illustrate in Fig. (c) and (d). The morphology of hydrothermal products are rod-like with diameters of ~52 nm (~43 nm) and length of less than 1μm for HTBFO (HTBLFO) nanoparticles. This morphological change via hydrothermal method, was strongly dependent on the synthesis process. It could be due to the nucleation rate of grains which is related to the condition of synthesis process. In both synthesis methods, The La concentration had only a slight effect on the morphology and only reduced the size of grains, which is in agreement with the results from Scherrer formula.

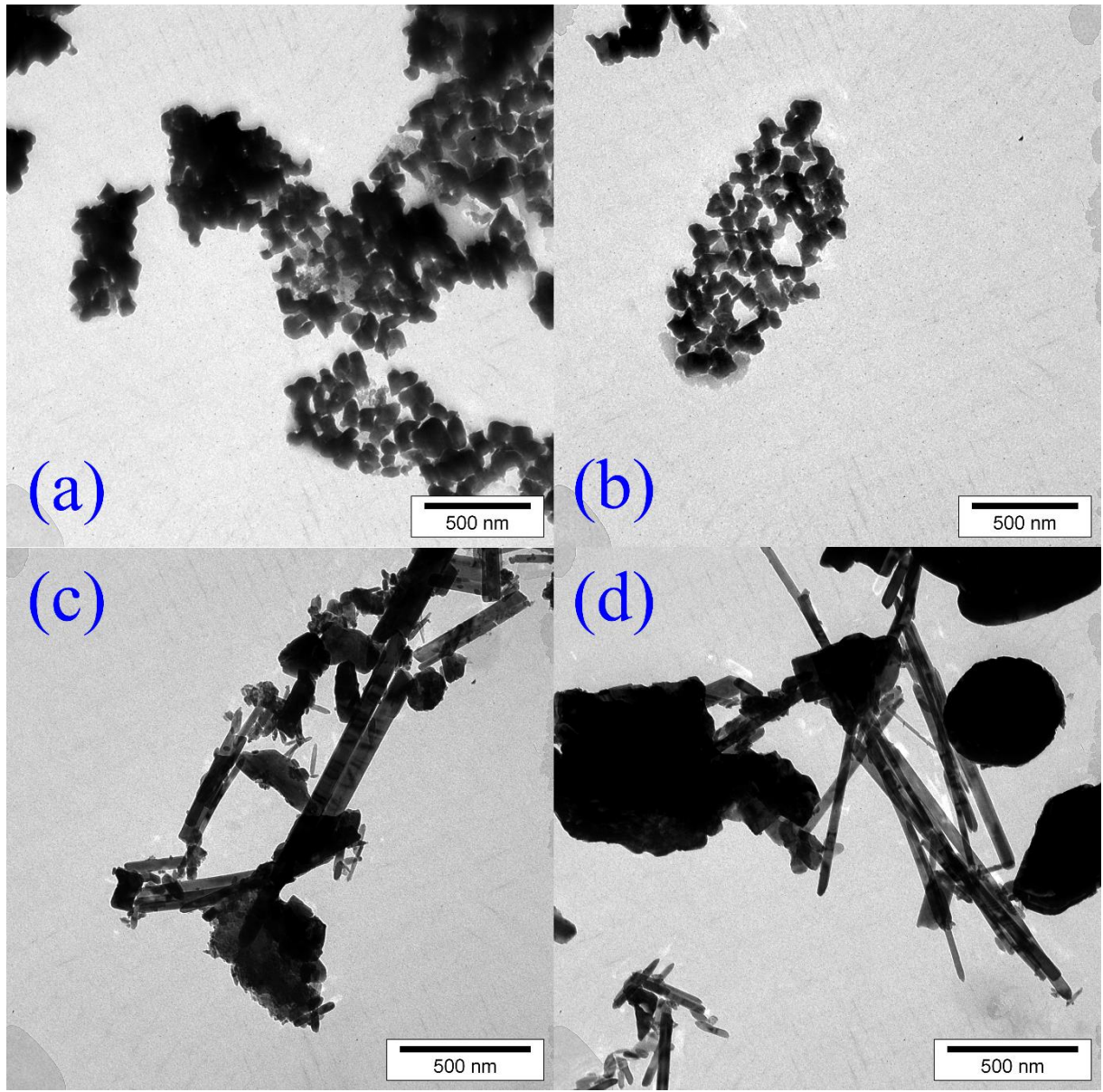

**Fig. 3.** TEM images of BiFeO$_3$ and Bi$_{0.8}$La$_{0.2}$FeO$_3$ synthesized by SG and HT methods, (a) image of SGBFO sample, (b) SGBLFO, (c) HTBFO and (d) HTBLFO

### 3.4 Thermal behavior

Fig. 4 shows the thermal gravitometric and differential thermal analysis (TG/DTA) curves of all products. In DTA curves, exothermic peaks are related to the crystallization and oxidation. On the other hand, phase transition and dehydration show endothermic peaks. TG curves of SGBFO and SGBLFO samples reveal the decomposition of the organic part of samples up to 400°C, with a wide exothermic peak on the DTA curves and ~1.3% of mass loss.

For the HTBFO (HTBLFO) samples (inset in Fig. 4 (b)), the total weight loss was 2.3% (1.5%) which is divided to three stages for HTBFO nanoparticles. First loss of 0.5% up to 300°C and then quick loss of 1% up to 390°C and finally 0.8% weight loss up to 700°C which is related to the complete decomposition of nanoparticles. When the lanthanum is added into the BFO structure, Curie temperature decreases from 830°C (827°C) for SGBFO (HTBFO) to 824°C (818°C) for SGBLFO (HTBLFO). The reason could be explained by lower polarizability of lanthanum ions in the structure of bismuth ferrite, compared to bismuth, as a result, the Curie temperature decreases [52].

**3.5 M-H hysteresis loops analysis of BFO and BNLFO nanoceramics**

The magnetic hysteresis loops of as-prepared samples were measured using VSM at RT with a maximum applied field of 20 KG (Fig.5). For all products a weak ferromagnetic behavior is observed. However HTBFO and HTBLFO nanoparticles didn't show a completed saturation magnetization. A significant enhancement in the value of saturation magnetization ($M_s$) on SG synthesis method is observed due to the smaller particle size and morphology. The M-H hysteresis loops data (Table 1), indicated that the remnant magnetization ($M_r$) and $M_s$ decreases with addition of lanthanum. However the coercive forces increase. Dopant La and changes in the distribution of the bismuth and consequently iron ions in the A-site and B-site of BFO structure, changes the magnetic moments and anisotropy [16,26]

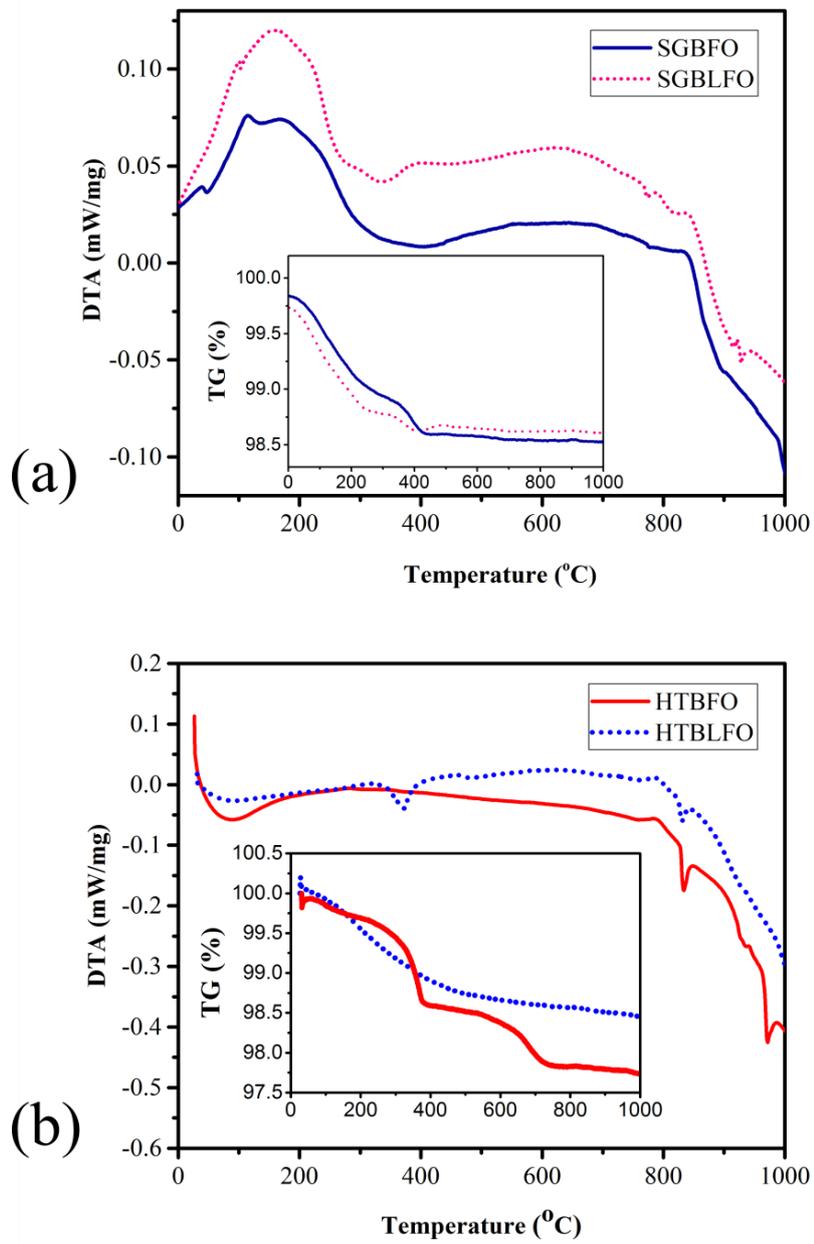

**Fig. 4** (a) DTA curves of pure and 20%La-doped bismuth ferrite nanoparticles prepared by SG method. The inset shows the TG curves of products, (b) DTA and TGA (as inset) curves of HTBFO and HTBLFO nanorods.

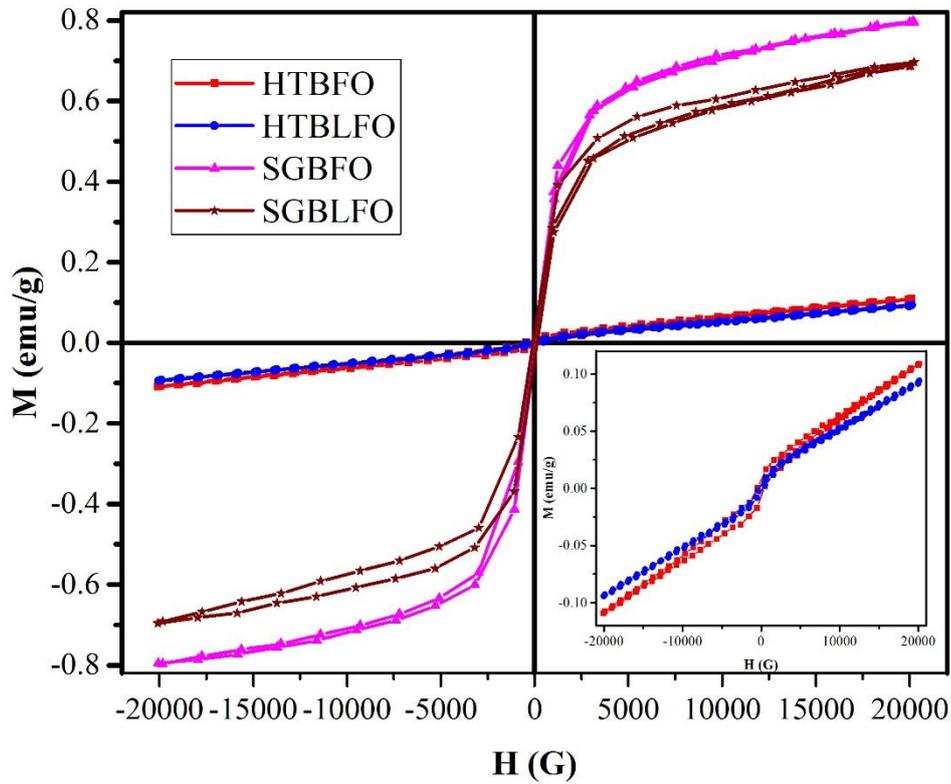

**Fig. 5** M-H hysteresis loops of all as-prepared samples through SG and HT methods at RT. Inset shows hysteresis loops of $BiFeO_3$ and $Bi_{0.8}La_{0.2}FeO_3$ nanoparticles prepared by HT method separately

Table 1. The saturation and remanent magnetization, as well as coercively of BFO and BLFO powders synthesized by SG and HT methods.

| Sample | Sol gel | | | Hydrothermal | | |
|---|---|---|---|---|---|---|
| | $M_s$(emu/g) | $M_r$(emu/g) | $H_c$(G) | $M_s$(emu/g) | $M_r$(emu/g) | $H_c$(G) |
| $BiFeO_3$ | 0.797 | $15.596\times10^{-3}$ | 39.034 | 0.109 | $7.077\times10^{-3}$ | 383.653 |
| $Bi_{0.8}La_{0.2}FeO_3$ | 0.696 | $10.452\times10^{-3}$ | 90.788 | 0.094 | $2.681\times10^{-3}$ | 426.832 |

### 3.6 Uv-vis spectroscopy analysis

The optical spectra of as-prepared products as shown in Fig. 6 reveals a high absorption in the range of 500-600 nm for all samples. The direct optical band-gap energy of all nanoparticles was calculated using Tauc's equation $(\alpha h\nu)^2=K(h\nu-E_g)$, where K is a constant, α is the absorption coefficient and *hν* is the photon energy [53]. $E_g$ is determined by plotting $(\alpha h\nu)^2$ vs hν and extrapolation of the straight line portion of the curves. The band gap decreases by La-substitution into the BFO structure. The optical band-gap was calculated 2.12eV (2.13 eV) for SGBLFO (HTBLFO) nanoparticles which was lower than pure BFO samples (2.16 eV for SGBFO and 2.17 eV for HTBFO).

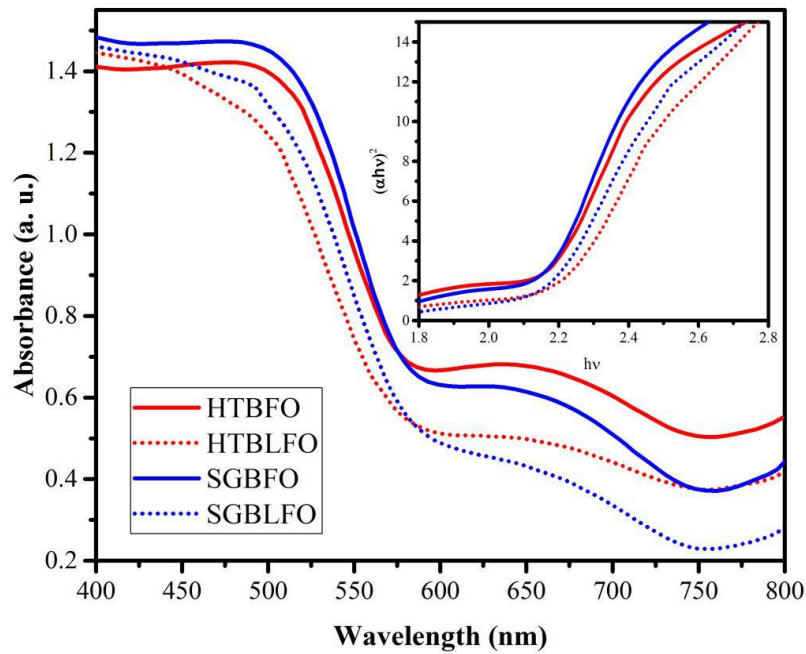

**Fig. 6** UV-vis adsorption of of BiFeO$_3$ and Bi$_{0.8}$La$_{0.2}$FeO$_3$ nanoparticles prepared by both SG and HT methods

### 3.7 Photocatalytic activity of BFO and BNFO

The photocatalytic activity of BFO sand BLFO nanoparticles is evaluated by the degradation of methylene blue (MB) under visible light irradiation. Herein, 0.2g sample was mixed by 4 mg/L of MB solution at RT. The photocatalytic degradation is measured from relation $C/C_0$, where $C$ and $C_0$ are concentration at time $t$ and zero respectively considering the time unit minutes. Fig. 7 (a) shows the photo-gradation coefficients of all products vs. time. For SGBFO and HTBFO grains, 26% and 21% of degradation take place in 300 min. However 71% and 58% of degradation take place with SGBLFO and SGBFO nanocatalysts in 5h respectively. In principle, the efficiency of the photocatalysts depends on the factors like crystallite size, morphology, surface area, and band gap and photoinduced electron-hole separation efficiency of the catalyst. Fig. 7 (b) shows the rate constant of $\ln(C_0/C)$ vs. time. The rate constant k is calculated from the slopes of the fitted kinetics ($\ln(C/C_o) = -kt$) [54].

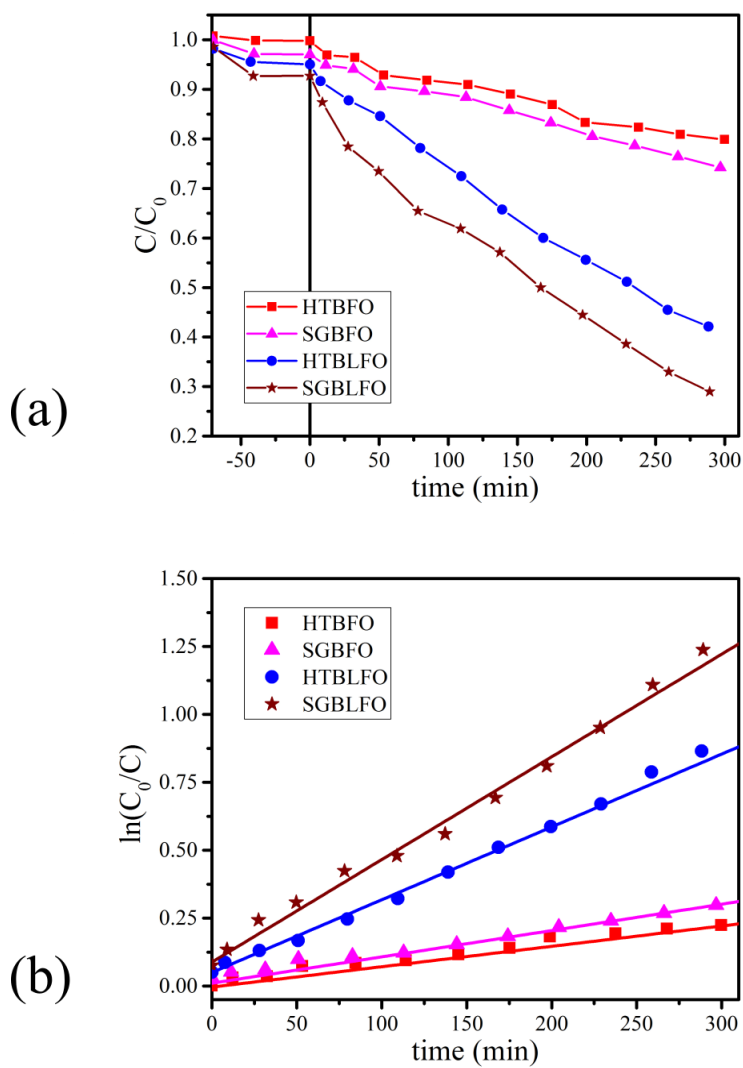

**Fig. 7** (a) Degradation of MB solution in the presence of sol-gel and hydrothermal prepared BiFeO$_3$ and Bi$_{0.8}$La$_{0.2}$FeO$_3$ nanoparticles, (b) first order kinetics data for degradation of MB on SGBFO, SGBLFO, HTBFO and HTBLFO.

## 4. Conclusions

In summary, pure and La-doped BiFeO$_3$ nanoparticles were synthesized by two different sol-gel and hydrothermal methods. The effect of synthesis method, as well as, the enhancement of photocatalytic activity of bismuth ferrite nanoparticles by addition of lanthanum is investigated. XRD analysis confirms the rhombohedrally-distorted (R$_3$C) perovskite structure for all products. The TEM studies reveal that as-prepared nanoparticles synthesized by SG are semi-spherical powders. On the other hand, as-prepared samples synthesized by HT have rod-like shape. DTA study indicated that addition of lanthanum in the Bi-site in the structure of BFO decreases the Curie temperature ($T_C$). Magnetic hysteresis loops showed the influence of synthesis method on the magnetic properties of samples. SGBFO and SGBLFO samples have much higher saturated magnetization which could be related to destruction of the spin cycloid magnetic ordering structure compared to rod-like HTBFO and BLFO nanoparticles. The photocatalytic reactivity of all samples was evaluated in terms of the degradation of MB in aqueous BFO and BLFO suspensions under visible light irradiation. The results indicated that the photocatalytic activity of products depends not only on the synthesis method but also on the addition lanthanum into the structure of bismuth ferrite.

## Acknowledgments

This work was supported by the research council of Shahid Bahonar University of Kerman (SBUK-RC). Partial financial support by Iranian nanotechnology initiative council (INIC) are acknowledged.

## 5. References

[1] M. Humayun, A. Zada, Z. Li, M. Xie, X. Zhang, Y. Qu, F. Raziq, L. Jing, Applied Catalysis B: Environmental 180 (2016) 219.

[2] Y. Wu, W. Han, S.X. Zhou, M.V. Lototsky, J.K. Solberg, V.A. Yartys, Journal of Alloys and Compounds 466(1–2) (2008) 176.

[3] S. Wang, D. Chen, F. Niu, N. Zhang, L. Qin, Y. Huang, Journal of Alloys and Compounds 688 (2016) 399.

[4] C.M. Cho, J.H. Noh, I.-S. Cho, J.-S. An, K.S. Hong, J.Y. Kim, Journal of the American Ceramic Society 91(11) (2008) 3753.

[5] F. Gao, X.Y. Chen, K.B. Yin, S. Dong, Z.F. Ren, F. Yuan, T. Yu, Z.G. Zou, J.-M. Liu, Advanced Materials 19(19) (2007) 2889.

[6] Y.-L. Pei, C. Zhang, Journal of Alloys and Compounds 570 (2013) 57.

[7] H. Maleki, Journal of Magnetism and Magnetic Materials 458 (2018) 277.

[8] J. Wang, J.B. Neaton, H. Zheng, V. Nagarajan, S.B. Ogale, B. Liu, D. Viehland, V. Vaithyanathan, D.G. Schlom, U. V Waghmare, N.A. Spaldin, K.M. Rabe, M. Wuttig, R. Ramesh, Science (New York, N.Y.) 299(5613) (2003) 1719.

[9] W. Eerenstein, N.D. Mathur, J.F. Scott, Nature 442(7104) (2006) 759.

[10] J. Ma, J. Hu, Z. Li, C.-W. Nan, Advanced Materials 23(9) (2011) 1062.

[11] H. Maleki, M. Haselpour, R. Fathi, Journal of Materials Science: Materials in Electronics 29(5) (2018) 4320.

[12] A. Sarkar, G.G. Khan, A. Chaudhuri, A. Das, K. Mandal, Applied Physics Letters 108(3) (2016) 33112.

[13] S.K. Mandal, T. Rakshit, S.K. Ray, S.K. Mishra, P.S.R. Krishna, A. Chandra, Journal of Physics: Condensed Matter 25(5) (2013) 55303.

[14] Y. Bai, T. Siponkoski, J. Peräntie, H. Jantunen, J. Juuti, Applied Physics Letters 110(6) (2017) 63903.

[15] P.P. Biswas, T. Chinthakuntla, D. Duraisamy, G. Nambi Venkatesan, S. Venkatachalam, P. Murugavel, Applied Physics Letters 110(19) (2017) 192906.

[16] G. Catalan, J.F. Scott, Advanced Materials 21(24) (2009) 2463.

[17] D. Wang, M. Wang, F. Liu, Y. Cui, Q. Zhao, H. Sun, H. Jin, M. Cao, Ceramics International 41(7) (2015) 8768.

[18] H. Fki, M. Koubaa, L. Sicard, W. Cheikhrouhou-Koubaa, A. Cheikhrouhou, S. Ammar-Merah, Ceramics International 43(5) (2017) 4139.

[19] G.A. Smolenskii, V.A. Bokov, Journal of Applied Physics 35(3) (1964) 915.

[20] P. Fischer, M. Polomska, I. Sosnowska, M. Szymanski, Journal of Physics C: Solid State Physics 13(10) (1980) 1931.

[21] I. Sosnowska, T.P. Neumaier, E. Steichele, Journal of Physics C: Solid State Physics 15(23) (1982) 4835.

[22] J.M. Moreau, C. Michel, R. Gerson, W.J. James, Journal of Physics and Chemistry of Solids 32(6) (1971) 1315.

[23] R.T. Smith, G.D. Achenbach, R. Gerson, W.J. James, Journal of Applied Physics 39(1) (1968) 70.

[24] C. Ederer, N.A. Spaldin, Physical Review B 71(6) (2005) 60401.

[25] D. Sando, A. Barthélémy, M. Bibes, Journal of Physics: Condensed Matter 26(47) (2014) 473201.

[26] J.T. Heron, D.G. Schlom, R. Ramesh, Applied Physics Reviews 1(2) (2014) 21303.

[27] F. Huang, X. Lu, W. Lin, X. Wu, Y. Kan, J. Zhu, Applied Physics Letters 89(24) (2006) 242914.

[28] Y.P. Wang, L. Zhou, M.F. Zhang, X.Y. Chen, J.-M. Liu, Z.G. Liu, Applied Physics Letters 84(10) (2004) 1731.

[29] J.-H. Xu, H. Ke, D.-C. Jia, W. Wang, Y. Zhou, Journal of Alloys and Compounds 472(1–2) (2009) 473.

[30] J.B. Neaton, C. Ederer, U. V. Waghmare, N.A. Spaldin, K.M. Rabe, Physical Review B 71(1) (2005) 14113.

[31] T. Zhao, A. Scholl, F. Zavaliche, K. Lee, M. Barry, A. Doran, M.P. Cruz, Y.H. Chu, C. Ederer, N.A. Spaldin, R.R. Das, D.M. Kim, S.H. Baek, C.B. Eom, R. Ramesh, Nature Materials 5(10) (2006) 823.

[32] D. Lebeugle, D. Colson, A. Forget, M. Viret, P. Bonville, J.F. Marucco, S. Fusil, Physical Review B 76(2) (2007) 24116.

[33] R. Safi, H. Shokrollahi, Progress in Solid State Chemistry 40(1–2) (2012) 6.

[34] J. Silva, A. Reyes, H. Esparza, H. Camacho, L. Fuentes, Integrated Ferroelectrics 126(1) (2011) 47.

[35] D. Varshney, A. Kumar, K. Verma, Journal of Alloys and Compounds 509(33) (2011) 8421.

[36] H. Singh, K.L. Yadav, Journal of Physics: Condensed Matter 23(38) (2011) 385901.

[37] Y.-H. Lee, J.-M. Wu, C.-H. Lai, Applied Physics Letters 88(4) (2006) 42903.

[38] F.Z. Li, H.W. Zheng, M.S. Zhu, X.A. Zhang, G.L. Yuan, Z.S. Xie, X.H. Li, G.T. Yue, W.F. Zhang, J. Mater. Chem. C 5(40) (2017) 10615.

[39] T. Durga Rao, S. Asthana, Journal of Applied Physics 116(16) (2014) 164102.

[40] X. Qi, J. Dho, R. Tomov, M.G. Blamire, J.L. MacManus-Driscoll, Applied Physics Letters 86(6) (2005) 62903.

[41] X. Zhang, Y. Sui, X. Wang, Y. Wang, Z. Wang, Journal of Alloys and Compounds 507(1) (2010) 157.

[42] Q.R. Yao, Y.H. Shen, P.C. Yang, H.Y. Zhou, G.H. Rao, Z.M. Wang, J.Q. Deng, Ceramics International 42(5) (2016) 6100.

[43] B. Yotburut, P. Thongbai, T. Yamwong, S. Maensiri, Ceramics International 43(7) (2017) 5616.

[44] M. Rangi, S. Sanghi, S. Jangra, K. Kaswan, S. Khasa, A. Agarwal, Ceramics International 43(15) (2017) 12095.

[45] R. Guo, L. Fang, W. Dong, F. Zheng, M. Shen, The Journal of Physical Chemistry C 114(49) (2010) 21390.

[46] Y. Zhang, A.M. Schultz, P.A. Salvador, G.S. Rohrer, Journal of Materials Chemistry 21(12) (2011) 4168.

[47] H. Maleki, S. Zare, R. Fathi, Journal of Superconductivity and Novel Magnetism (2017) 1.

[48] H. Maleki, S. Falahatnezhad, M. Taraz, Journal of Superconductivity and Novel Magnetism (2018) 1.

[49] A. Brisdon, Applied Organometallic Chemistry 24(6) (2010) n/a.

[50] T. Gholam, A. Ablat, M. Mamat, R. Wu, A. Aimidula, M.A. Bake, L. Zheng, J. Wang, H. Qian, R. Wu, K. Ibrahim, Journal of Alloys and Compounds 710 (2017) 843.

[51] X. Yang, G. Xu, Z. Ren, X. Wei, C. Chao, S. Gong, G. Shen, G. Han, CrystEngComm 16(20) (2014) 4176.

[52] S. Karimi, I.M. Reaney, I. Levin, I. Sterianou, Applied Physics Letters 94(11) (2009) 112903.

[53] J. Tauc, Materials Research Bulletin 5(8) (1970) 721.

[54] † X. H. Wang, † J.-G. Li, †,‡ H. Kamiyama, ‡ and Y. Moriyoshi, † T. Ishigaki*, (2006).